\documentclass{aa}  

\usepackage{graphicx}
\usepackage{txfonts}
\begin{document}

   \title{Beryllium enhancement in  stars of  the accreted Thamnos-2 system}


   \author{P. Molaro
          \inst{1,2}
          \and
          P. Bonifacio \inst{1,3}
          \and
          E. Caffau \inst{1,3}
          \and
          L. Monaco \inst{1,4}
          \and
          G. Cescutti \inst{1,2,5,6}
          }

   \institute{INAF-OATs, Via G.B. Tiepolo 11, I 34143 Trieste, Italy
         \and
         Institute of Fundamental Physics of the Universe. Via Beirut 2, I34151, Trieste Italy
         \and
             LIRA, Observatoire de Paris , Universit\'e PSL, Sorbonne Universit\'e, Universit\'e Paris Cit\'e, CY Cergy Paris Universit\'e, CNRS, 92190 Meudon, France 
             \and
             Facultad de Ciencias Exactas, Departamento de Física y Astronomía- Instituto de Astrofísica, Universidad Andres Bello, Autopista Concepción-Talcahuano, Talcahuano, 7100, Chile
             \and
             Department of Physics, Astronomy Section,University of Trieste, Via G. B. Tiepolo 11, Trieste, 34143, Italy
             \and
             INFN, Sezione di Trieste, via Valerio 2, I-34134 Trieste, Italy.
             }

   \date{Received September 15, 1996; accepted March 16, 1997}

  \abstract
  { Surveys of Galactic halo stars have revealed numerous streams and substructures tracing stellar populations accreted by the Milky Way. Among these, Gaia-Sausage-Enceladus (GSE) and Sequoia are the most prominent, both associated with dwarf galaxies accreted about 10 Gyr ago. }  
   {We aim to measure beryllium abundances in nine stars associated with Thamnos, a substructure possibly linked to Sequoia, following the discovery of a Be-rich star BPM\,3066 by \citet{Monaco2025A&A...697A..75M}.  }
   { We used Gaia photometry and parallaxes to compute ATLAS9 model atmospheres. Synthetic spectra were generated with \texttt{turbospectrum} and used with MyGIsFOS in single-model mode to analyze UVES high resolution spectra.} 
   { Four new stars exhibit a significant beryllium overabundance. Moreover, the two known Be-rich stars, HD 106038 and HD 132475, are also found consistent with Thamnos membership. Thus, all currently known Be-rich stars  appear associated with the Thamnos-2 structure. The Be enhancement is accompanied by elevated Si abundances, and we detect a correlation between Be and neutron-capture elements. No comparable Be-rich population is known elsewhere in the Galaxy, pointing to a rare enrichment event.}
   {
   The measured A(Be)/A(Li) excess ratio bears the imprint of spallation reactions, pointing to a highly energetic event in which fast CNO nuclei fragmented upon collision with the surrounding medium.  The silicon overabundance is also consistent with a hypernova origin. Such an event may have rapidly enriched the surrounding gas to [Fe/H]$\simeq-1.5$, explaining the relatively high metallicities of stars formed from this material despite their old ages ($\approx$ 13 Gyr).
    }

   \maketitle

\section{Introduction}

Studies of the Galactic halo have uncovered a wealth of stellar streams and substructures, remnants of past accretion events. Notably, Gaia-Sausage-Enceladus (GSE) and Sequoia stand out as prominent examples, representing dwarf galaxies that were disrupted and merged with the Milky Way approximately 10 Gyr ago
\citep{Belokurov2018,Helmi2018,Haywood2018,myeong2019MNRAS.488.1235M}.

The stellar debris of such mergers provides a unique opportunity to study, in situ, the chemical properties of extragalactic systems, in particular of trace elements, such as $^{9}$Be,  that cannot be measured in stars of external galaxies, even with next-generation facilities such as the E-ELTs. $^{9}$Be is produced exclusively through spallation of CNO nuclei by energetic cosmic rays, or through collisions of energetic CNO nuclei with protons and $\alpha$-particles in the interstellar medium \citep{Reeves1970}. Since both the cosmic-ray flux and CNO abundances may differ across galaxies, distinct Be enrichment histories could be possible. Indeed, \citet{molaro2020} reported that the [Be/H]–[Fe/H] relation in GSE stars is less scattered and holds a different slope compared to that of the Milky Way.

Lithium provides an additional, complementary case. Unlike Be, Li has multiple production channels, including primordial nucleosynthesis, spallation, AGB stars, and novae. Yet the Li abundances observed in unevolved halo stars fall below the present value, implying subsequent Galactic enrichment. \citet{molaro2020} found that Li abundances in low-metallicity GSE candidates overlap with the Galactic halo plateau (A(Li) = 2.18), while \citet{simpson2021MNRAS.507...43S} reported a slightly higher mean value (A(Li) = 2.40).

The Sequoia system offers another  environment in which to probe light-element enrichment. Anomalously high Be abundances — higher than  the solar value — have been reported in one of its stars \citep{Monaco2025A&A...697A..75M}. Explanations such as planetary engulfment or hypernova-driven enrichment have been proposed but remain unconstrained. In this work, we extend the analysis to nine additional Sequoia members, which all belong to the low energy component more recently identified as  Thamnos  substructure \citep{koppelman2019A&A...631L...9K,feuillet2021MNRAS.508.1489F,dodd2023A&A...670L...2D}.  A chemical analysis of  Thamnos candidates have revealed a high level of  contamination from the in situ population \citep{ceccarelli2025A&A...704A.180C}.
However, most of the stars examined in this study also exhibit a Be enhancement, a feature not observed elsewhere in the Galaxy, thereby offering a powerful means of chemical tagging for membership identification.

\section{Target  selection}\label{secA1}

Candidate stars were identified from the GALAH value-added catalog, which includes dynamical information, adopting the selection criteria of \citet{feuillet2021MNRAS.508.1489F}. Additional constraints of $T_{\rm eff} > 5700$ K and $\log g > 3.65$ were imposed to exclude stars affected by beryllium depletion or dilution during main-sequence or post-main-sequence evolution. We further required high-quality astrometry (RUWE $<$ 1.4 and parallax/error $>$ 5) and reliable GALAH parameters flags (flag\_sp = 0, flag\_fe\_h = 0), resulting in a sample of 38 stars. From this set, the brightest targets ($G < 13$) were selected to ensure that Be abundances could be measured with reasonable exposure times. Twelve stars were observed, two of which were later identified as spectroscopic binaries and are excluded from this work, leaving a final sample of 10 stars. One object, BPM\,3066, has already been analysed by \citet{Monaco2025A&A...697A..75M}  using in the analysis the same methodology followed here. The Be II 313.0 nm region was observed using the UVES spectrograph (blue arm, DIC1 346+580 nm setting). Basic stellar parameters and some details of the observations  are listed in Table \ref{table1}.

For convenience, the targets are labeled with the prefix TS (Thamnos Star) followed by a sequential number. We also include two well-studied comparison stars, HD 132475 and HD 106038, which were reanalysed using archival UVES data for homogeneity. A posteriori inspection of the orbital properties in the ($E$–$L_z$) plane confirms that all selected stars are consistent with membership in the Thamnos structure.

\begin{figure}
\centering
 \resizebox{0.45\textwidth}{!}{\includegraphics[clip=true]{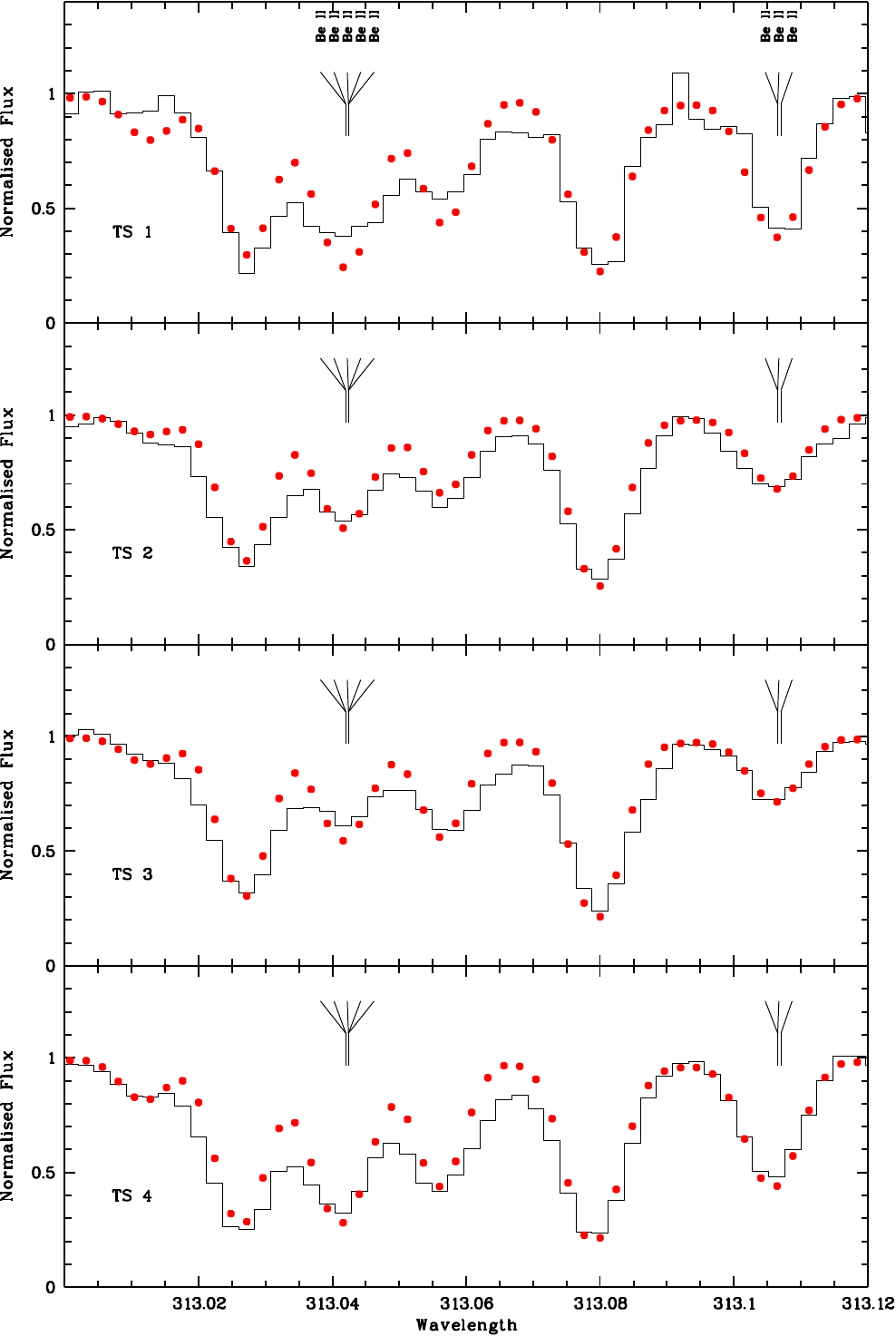}}
    \caption{ Spectral regions for the program stars around the BeII 313.0 nm doublet lines together with the best fit (red dots)  for stars TS\,1, TS\,2, TS\,3 and TS\,4. }
    \label{fig:fits}
\end{figure}

\begin{figure}
\centering
 \resizebox{0.5\textwidth}{!}{\includegraphics[clip=true]{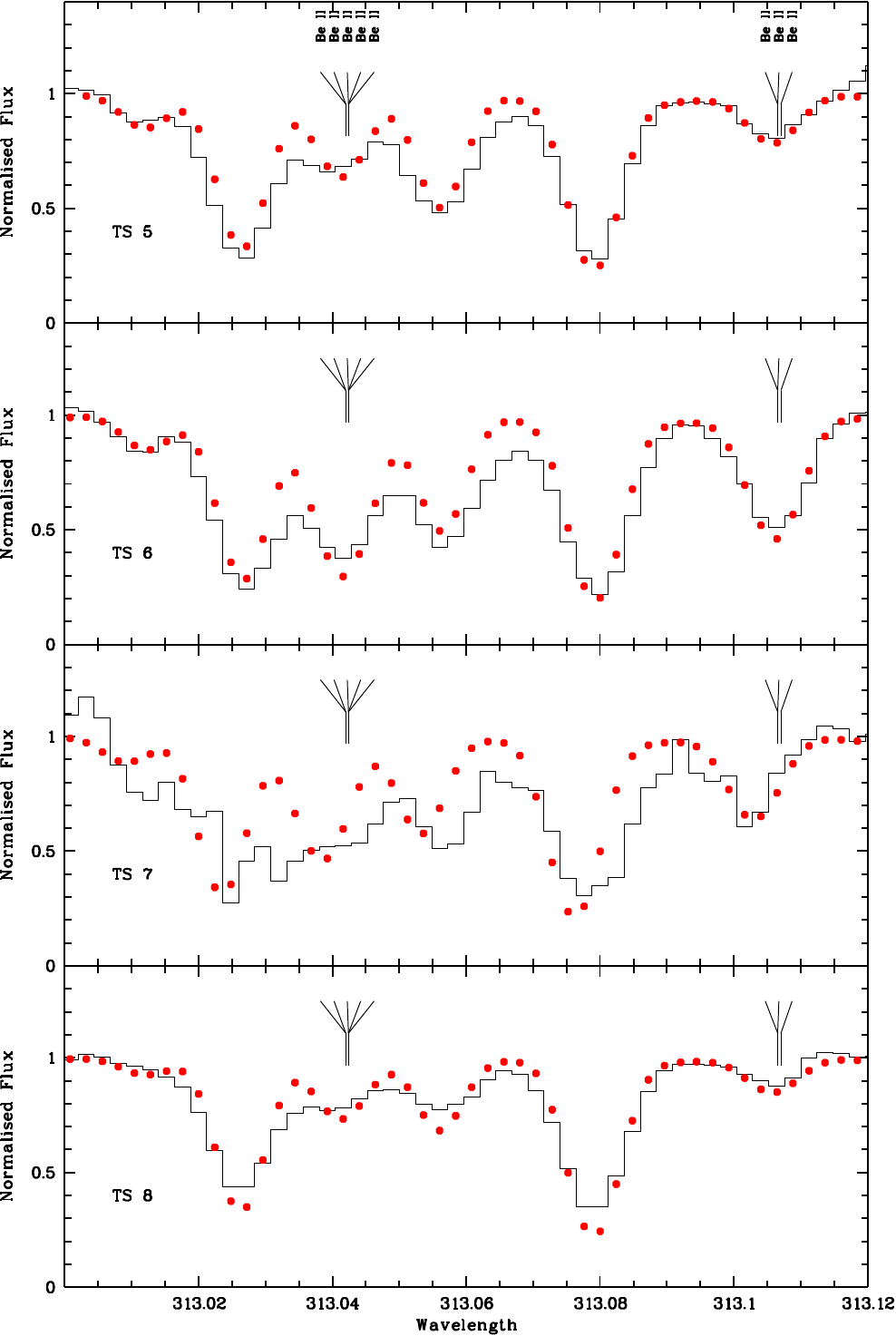}}
    \caption{ As in Fig \ref{fig:fits} for stars TS\,5, TS\,6, TS\,7 and TS\,8}
    \label{fig:fits1}
\end{figure}

\begin{figure}
\centering
 \resizebox{0.5\textwidth}{!}{\includegraphics[clip=true]{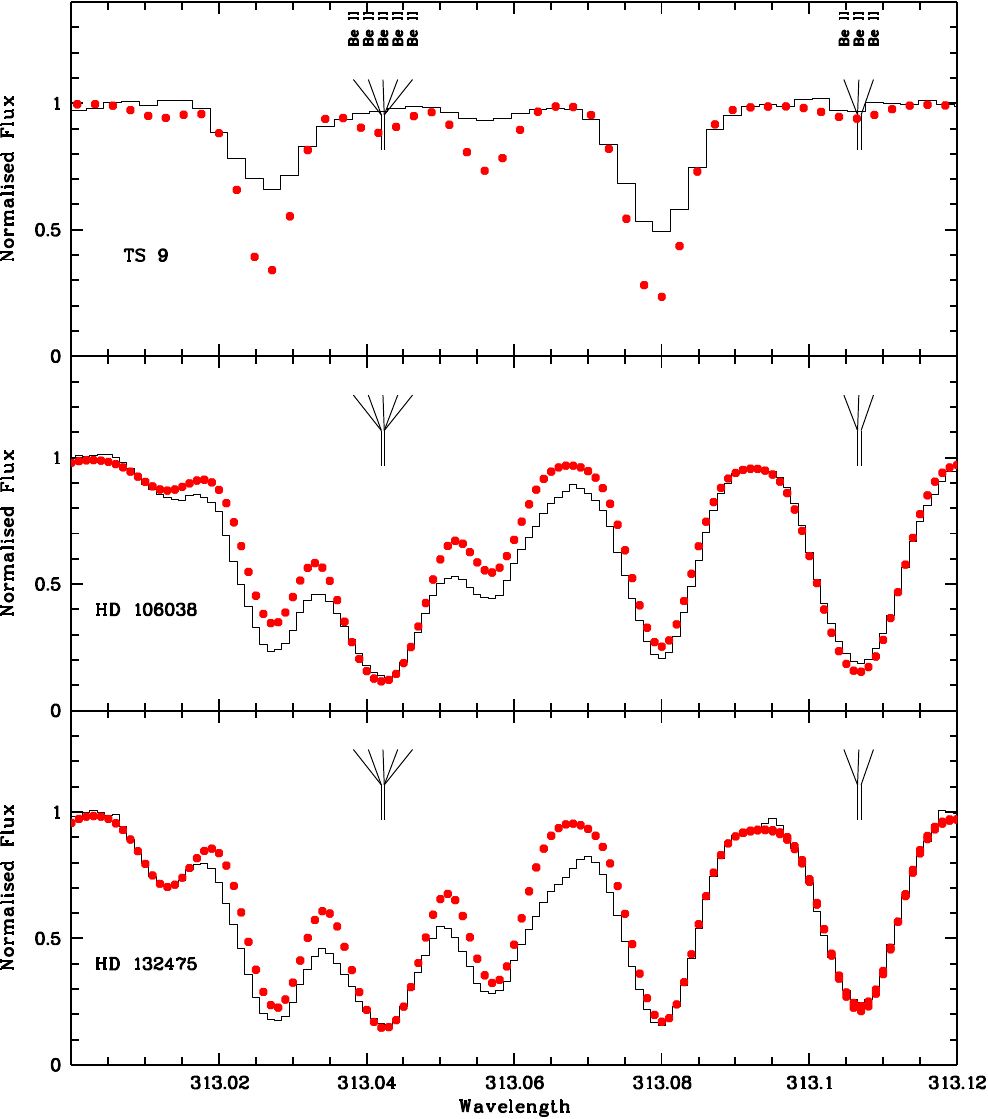}}
    \caption{ As in Fig \ref{fig:fits}  for stars TS\,9, HD\,106038 and HD\,132475.  }
    \label{fig:fits2}
\end{figure}

\begin{figure}
 \includegraphics[width=0.5\textwidth]{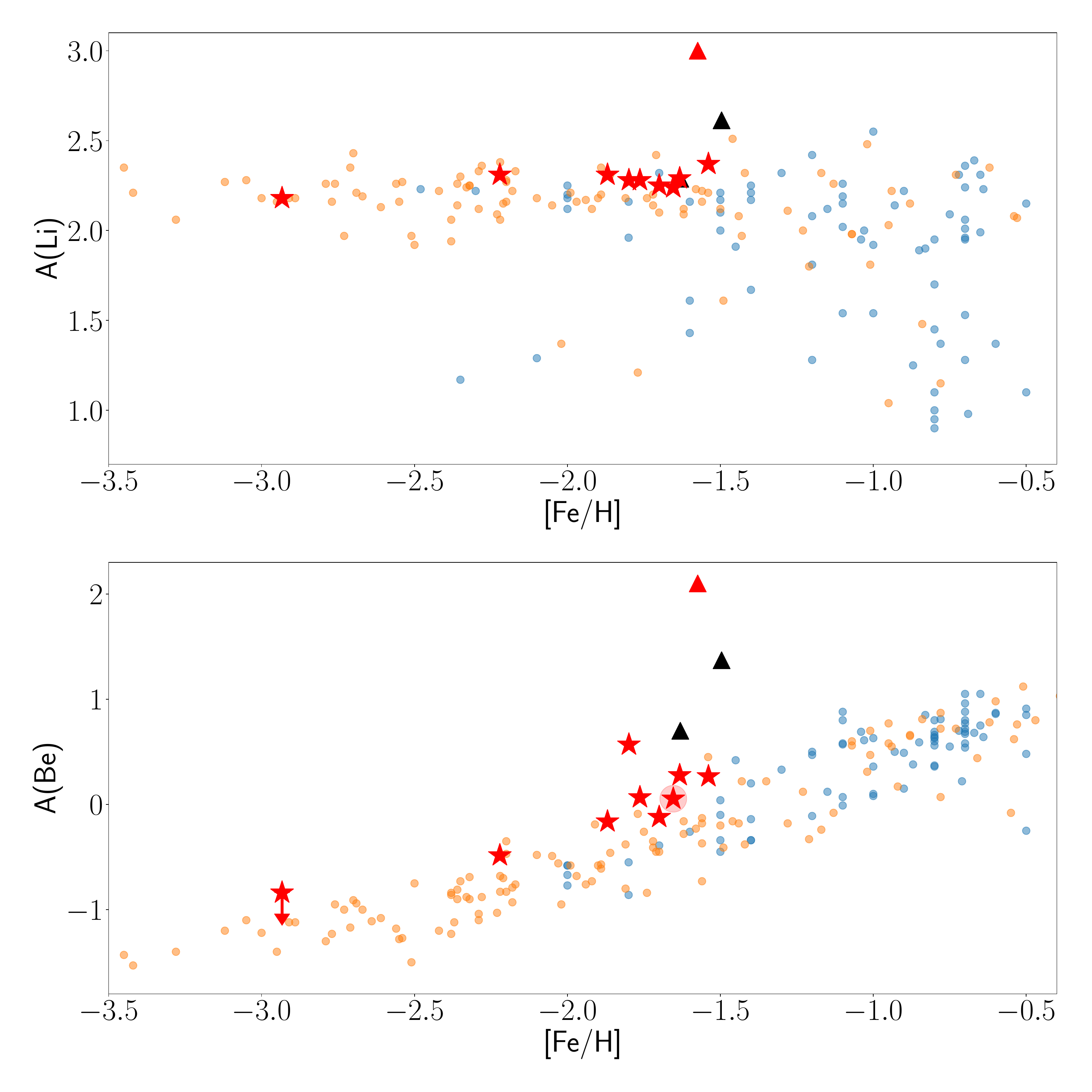}
    \caption{ 
    Top panel: Li abundances;  Bottom panel: beryllium abundances.    Stars from Thamnos  are shown in red.  HD 106038, HD 132475, and BPM\,3066  sharing the   kinematical properties are also shown as back and red triangles. Background Li and Be   are 
 from the compilations of  \citet{Boesgaard2011},  and of \citet{Smiljanic2009}, with orange and light blue colour dots, respectively.   }
    \label{fig:fig_libe}
\end{figure}

\begin{figure}
 \includegraphics[width=0.5\textwidth]{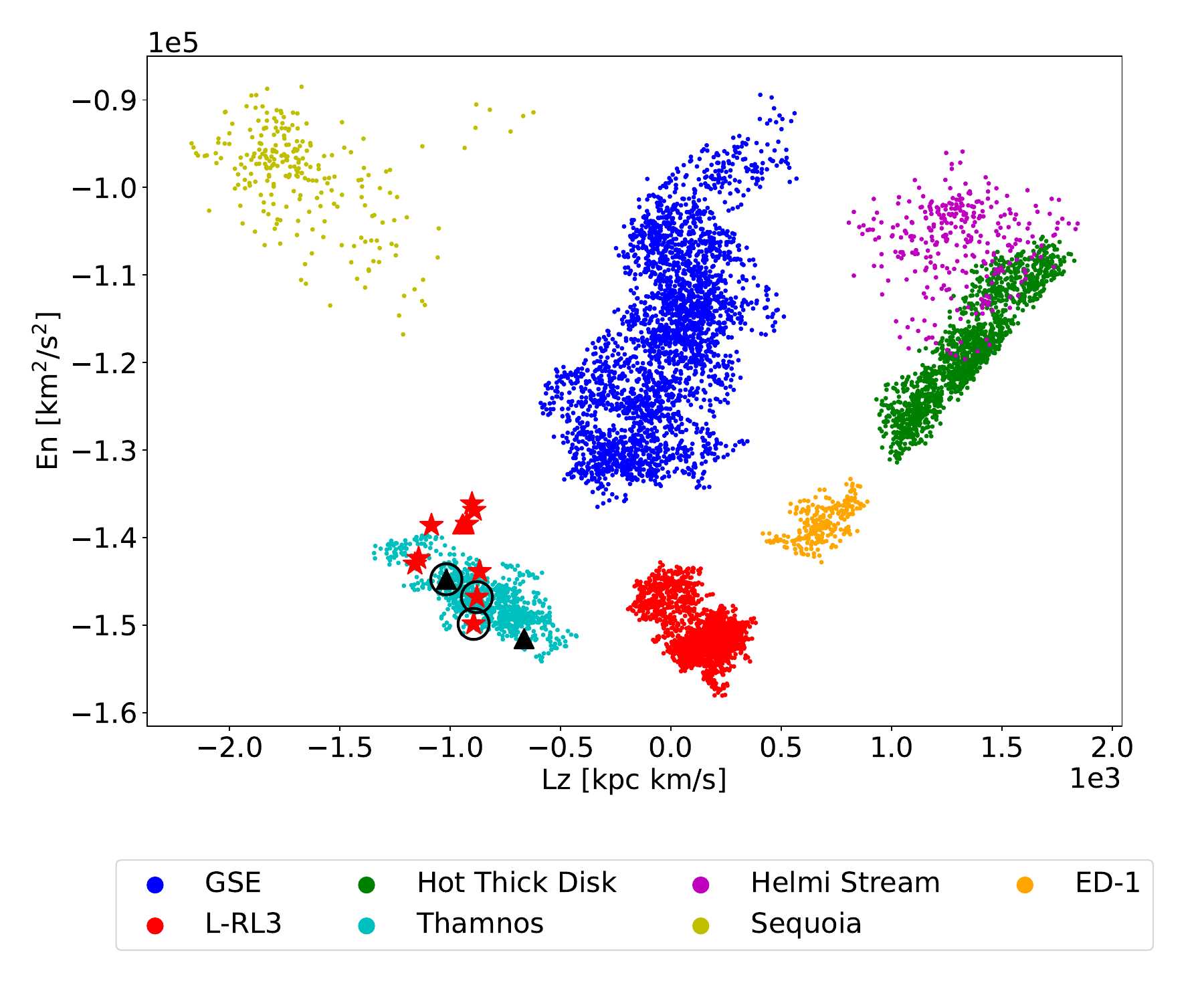}
    \caption{ Galactic halo stars groups according to  \citet{dodd2023A&A...670L...2D}  with seven main groups indicated in different colours.  The light green corresponds to Sequoia. The green is the hot thick disk  and the Helmi Stream is magenta. The ED-1 (Emma Dodd 1) is  orange. The GSE is in blue and the Lodal Ruiz-Lara 3 (L-RL3) in red. The cyan group  near which most of our stars are located corresponds to Thamnos. Here are located also BPM\,3066, a red triangle, and  HD\,106038, and HD\,132475 shown as black triangles. The circled symbols identify  the stars that are classified as bona fide Thamnos members by  \citet{dodd2023A&A...670L...2D}}
    \label{fig:fig3}
\end{figure}

\section{Observations}

Observations of candidate Sequoia–Thamnos stars were obtained with the UVES spectrograph at the ESO Very Large Telescope (VLT) under programme 111.24HT (P.I. L. Monaco). Spectra covering the Be\,II line at 313.0 nm were acquired using the blue arm with a central wavelength of 346 nm. The DIC1 346+580 UVES setting was adopted. The red arm simultaneously provided high-quality spectra over a broad wavelength range, enabling a detailed chemical abundance analysis of the target stars.

The CCD was operated with $2 \times 2$ on-chip binning. The slit width was set to 1.0 arcsec in the red arm and 1.2 arcsec in the blue arm, yielding resolving powers of $R \sim 40{,}000$ and $R \sim 35{,}000$, respectively. Red-arm exposures were split into multiple sub-exposures to avoid saturation.

For the two Thamnos stars HD 106038 and HD 132475, we analysed archival UVES spectra retrieved from the ESO archive. For HD 106038, we used three blue-arm spectra obtained with the DIC1 setup centred at 346 nm and a slit width of 1.0 arcsec with $1 \times 1$ binning. Two exposures were acquired on April 12, 2000, as part of programme 65.L-0507, each with an integration time of one hour. The third spectrum was obtained on March 28, 2004, under programme 072.B-0585, with an exposure time of 3198 s. The three ESO-reduced spectra were co-added for the analysis. For the red arm, we used the DIC1 580 nm spectrum obtained simultaneously with the 346 nm observation in programme 072.B-0585.

For HD 132475, we used a blue-arm DIC1 spectrum centred at 346 nm obtained on April 12, 2000 (programme 65.L-0507), with a slit width of 1.0 arcsec and an integration time of 1200 s. For the red arm, we used a DIC1 580 nm spectrum obtained on September 7, 2003 (programme 71.B-0529), with a slit width of 1.0 arcsec and an integration time of 960 s.

\begin{table*}[t!]  
\vspace*{0mm}
 \caption{ Parameters of the program stars and observational details.   } 
 \label{table1}
\begin{tabular}{llccccrrcc}
 \hline
  name  &other  &  Gaia (DR3)       &Teff & log g & [Fe/H] & G& Texp Blue &SNR& \\
  &  &           & K &  &  & mag& sec &sec &\\
  \hline
  TS\,0 & BPM\,3066& 4667364088963367808    &  5910 &4.29 &-1.57 & 12.87& 593 &15& \\
  TS\,1   & &   4629330504289973888 &5939 &4.28 &-1.80& 12.54& 1546& 13&\\
  TS\,2  &  &  4881321114629997312 &  6181 &4.18 &-1.76& 12.88&5994 &17&\\
  TS\,3  & &   4881378632832221312 &6061 &3.98 &-1.87& 12.80 & 5994 &16 &\\
  TS\,4  & &   5677177901742938880 &5910 &3.94 &-1.6& 13.53&5994 &23 &\\
  TS\,5  &  &  5462136547157907200 &6063 &4.41 &-1.70 &12.59& 5994 &18&\\
  TS\,6  &  &  6162713951575758464 &5959 &3.93 &-1.54& 12.64&5994 &25 &\\
  TS\,7   & &   5781595596159463040 & 6089 &4.10 & -1.65& 12.45 & 773& 23&\\
  TS\,8   & &   4187454819959693440 &6195 &3.92 &-2.22&11.64&5994 &30&\\
  TS\,9  & &   6682809550247700096 &6308 &3.97 &-2.93&11.95&8991 &65&\\
  TS\,10 & HD132475  & 6232043867720079616   &  5733 &3.83 &-1.63 &8.39& 1200 &61&\\
  TS\,11 & HD106038  & 3920615854833203328   &  6060 &4.28 &-1.49 &10.02&10398&29&\\
\hline
\end{tabular}
\tablefoot{The exposure times correspond to the blue UVES setting, and the signal-to-noise ratio (S/N) is measured at the Be II resonance lines near 313.0 nm.}
\end{table*}

\begin{table}[t!]  
\vspace*{0mm}
 \caption{ Ages, and  A(Be) and A(Li) abundances  for the program stars.  } 
 \label{table2}
\begin{tabular}{lrlrlrrrl}
 \hline
 name   & Age& $\sigma$ &Be& $\sigma$&Li& Li &$\sigma$\\
    &          Gyrs&  &&&&$_{3DNLTE}$&\\
  \hline
  TS\,0 &   12.54& 1.04&2.10&0.20 &3.00 & 2.90 &0.09\\
  TS\,1 &  12.94& 0.79& 0.57&0.06&2.28&2.27& 0.01\\
  TS\,2 &  13.13&0.62&0.07&0.02&2.28&2.28&0.01\\
  TS\,3 & 13.52 & 0.27&-0.16&0.03&2.31&2.30&0.01\\
  TS\,4  &  13.53&0.25&0.28&0.02&2.29&2.29&0.01\\
  TS\,5 &  11.81& 1.72&-0.12&0.03&2.25&2.24&0.01\\
  TS\,6 &  13.44&0.34&0.27&0.02&2.37&2.37&0.01\\
  TS\,7  & 13.39&0.40&0.05&0.04&2.24&2.24&0.02\\
  TS\,8  &13.43&0.32&-0.48&0.04&2.31&2.31&0.01\\
  TS\,9  & 13.40&0.34&-0.84&0.11&2.18&2.19&0.01\\
  TS\,10 &13.38& 0.37 &0.70&0.01&2.29&2.28&0.01\\
  TS\,11 &  12.60&1.10&1.37&0.01&2.61&2.59&0.01\\
\hline
\end{tabular}
\tablefoot{ The results for BPM,3066 (TS,0) are taken from \citet{Monaco2025A&A...697A..75M}. Note that the quoted uncertainties do not include the contribution from uncertainties in the atmospheric parameters. }
\end{table}

\subsection{Chemical Abundances}\label{sec:chem_abund}

All spectra, both those acquired by us and those retrieved from the ESO archive, were analysed in a homogeneous manner. Stellar parameters were first determined using Gaia photometry and parallaxes, following the method described in \citet{lombardo21}. Starting from an initial estimate of the metallicity, we refined it using MyGIsFOS \citep{Sbordone2014} in its classical multi-model mode, iterating until convergence was achieved.

Model atmospheres were then computed with the ATLAS9 code \citep{K05}, adopting opacity distribution functions with a microturbulent velocity of 1 km s$^{-1}$ and the closest metallicity available in the KOALA database \citep{KOALA}. Using these atmospheres, we generated grids of synthetic spectra with \texttt{turbospectrum} \citep{plez2012,plez2025}, varying the abundances in steps of 0.2 dex. The spectra were subsequently analysed with MyGIsFOS in single-model mode, as described in \citet{caffau2024}.

Beryllium abundances were derived from the $^9$Be\,II resonance doublet at 313.0 nm, a spectral region close to the atmospheric cutoff and affected by significant blending, primarily with V\,II and Fe\,I lines. Owing to these complications, Be is among the most challenging elements to measure in Galactic halo stars \citep{Molaro1984,Molaro1997,Smiljanic2009,Boesgaard2011}. For both Be\,II and Li\,I lines, we performed direct line-profile fitting using our proprietary code \texttt{abbofit}, rather than relying on MyGIsFOS. 

Table \ref{tab:tababb} reports the derived abundances for TS1 as an example, while the results for the full sample are provided as an online supplementary CSV table.

As a consistency check, we reanalysed the well-studied stars HD~106038 and HD~132475.  For HD~106038, we derive A($^9$Be) = 1.37 $\pm$ 0.01, in excellent agreement with previous determinations \citep{tan2009,Smiljanic2009}. The lithium abundance, A($^7$Li) = 2.59 $\pm$ 0.01 (3D NLTE), is also consistent with literature values \citep{Asplund2006,tan2009}. For HD~132475, we obtain A($^9$Be) = 0.70 $\pm$ 0.01, in agreement with \citet{Boesgaard2011}.

\subsection{Be and Li abundances}\label{sec:be_li}

Stellar syntheses in the beryllium region for the eleven stars analysed in this work  are shown in Figures \ref{fig:fits}, \ref{fig:fits1}, and \ref{fig:fits2}. The analysis for  BPM\,3066 was given in \citet{Monaco2025A&A...697A..75M} where the same methodology was used. The derived lithium and beryllium abundances are reported in Tables \ref{table2} and \ref{tab:tababb}, and are displayed in Figure \ref{fig:fig_libe} together with data from Galactic surveys by \citet{Boesgaard2011} and \citet{smiljanic2008MNRAS.385L..93S}.

Five stars (TS1, TS2, TS4, TS6, and TS7) are clearly Be-rich, with enhancements of 0.4–0.7 dex above the linear Be–Fe relation \citep{Boesgaard2011,smiljanic2008MNRAS.385L..93S,molaro2020}. Two additional stars, TS3 and TS5, lie 0.25 and 0.28 dex above the relation, respectively; given the intrinsic scatter of $\approx 0.08$ dex, these may represent moderate enhancements. In contrast, TS8 and TS9 show Be abundances consistent with Galactic stars of similar metallicity. Notably, these two objects are also the most metal-poor in the sample ([Fe/H] = –2.2 and –2.9), whereas the remaining stars span the range –1.5 to –1.9.

In the broader Galactic context, such levels of Be enrichment are extremely rare. Among the $\sim$150 stars analysed by \citet{Boesgaard2011} and \citet{Smiljanic2009}, only two objects—HD~106038 and HD~132475—exhibit comparable abundances. Their metallicities ([Fe/H] = –1.3 and –1.49) closely match those of our Be-rich stars, suggesting that this phenomenon preferentially occurs in this metallicity regime. The most extreme case, BPM\,3066, reaches A($^9$Be) = 2.10, exceeding the solar value despite its low metallicity \citep{Monaco2025A&A...697A..75M}.

The kinematic properties of the sample are illustrated in Figure~\ref{fig:fig3}, which shows orbital energy ($E$) versus the vertical component of the angular momentum ($L_z$), following the classification of \citet{dodd2023A&A...670L...2D}. In this framework, Thamnos stars occupy the low-energy region and are characterised by mildly eccentric retrograde orbits \citep{koppelman2019A&A...631L...9K}. All program stars, together with BPM\,3066, lie in this region, consistent with membership in the Thamnos structure. Notably, HD~106038 and HD~132475 also fall in the same region of the diagram.

Following \citet{koppelman2019A&A...631L...9K} and \citet{ruiz2022A&A...665A..58R}, Thamnos can be divided into two components: Thamnos-1 ([Fe/H] $\approx -2.0$, $v_{\phi} \approx -200$ km s$^{-1}$) and Thamnos-2 ([Fe/H] $\approx -1.3$ to –1.5, $v_{\phi} \approx -150$ km s$^{-1}$). In this scheme, the two most metal-poor stars (TS8 and TS9), which show normal Be abundances, are likely associated with Thamnos-1, while the remaining eight stars, all Be-enhanced, belong to Thamnos-2. The enhanced beryllium abundances therefore provide additional support for a common origin within this structure.

Lithium abundances are shown in Figure \ref{fig:fig_libe}, together with comparison samples from \citet{Smiljanic2009} and \citet{Boesgaard2011}. The program stars follow the general trend of Galactic halo stars but tend to lie toward the upper envelope of the distribution, indicating marginally higher Li abundances. For the 10 stars with A(Li) $<$ 2.5, we obtain a weighted mean of A(Li) = 2.290 $\pm$ 0.003, compared to A(Li) = 2.199 $\pm$ 0.086 reported by \citet{Sbordone2010}. This slight enhancement may reflect a contribution from Li production via spallation processes associated with Be enrichment.

Despite its accreted origin, the Thamnos structure exhibits Li abundances consistent with those of Galactic halo stars. Similar behaviour has been observed in other accreted systems, such as GSE \citep{molaro2020,simpson2021MNRAS.507...43S} and the globular clusters $\omega$~Cen and M54 \citep{Monaco2010,Mucciarelli2014}. This suggests a universal Li abundance in metal-poor stars, which remains a factor of $\sim$3 below the value predicted by standard Big Bang nucleosynthesis.

\begin{table}
\caption{LTE abundances for TS1.    } 
\label{tab:tababb}
\begin{tabular}{lrrcrrcrc} 
\hline
\smallskip
       &  & & & TS\,1& && & \\
\hline
\smallskip
     X &   Sun  & NL&  A(X)  &  $\sigma$ &  [X/H]  &  [X/FeI]    \\
\hline
 LiI     &   3.28    & 1 &   2.28 & 0.01& -1.0  & 0.80&  \\
 LiI*     &       &      &   2.27 & 0.01& -& -&  \\
 BeII     &   1.38         & 2 &  0.57 &0.06&-0.81&0.99 &\\
 OH    &  8.76    &68&7.33&0.18& -1.43&& \\
 NaI  & 6.30     & 4&4.64&0.27&-1.66 &0.14&\\
 MgI  & 7.54     &3 &  6.21 &0.13 &-1.33&0.47&\\ 
 SiI  & 7.52      &6& 6.31 & 0.10&-1.41 &0.39&\\ 
 SiII & 7.52      &2 &  6.58&0.11&-0.95 &0.85&\\
 CaI  & 6.33     &20 &  5.00  &0.05 &-1.33 &0.47&\\ 
 ScI  & 3.10     & 1& 1.51&0.20&-1.59 &0.21\\
 ScII  & 3.10      &8 & 1.54&0.06&-1.51&0.29&\\
 TiI & 4.90    & 13&   3.50 &0.10& -1.41&0.39&  \\
 TiII  & 4.90     &26 &   3.61&0.10 &-1.29&0.51& \\
 VII  & 4.00      &2& 2.49&0.07&-1.51&0.17&\\
 CrI  & 5.64      & 7&    3.81&0.05&-1.84&-0.04&\\ 
 CrII  & 5.64      & 8&4.11&0.08&-1.53&0.27& \\
 MnI   & 5.37      & 4&3.28&0.17&-2.09&-0.29& \\
 MnII   & 5.37      &2 &3.44&0.03&-1.93&-1.31&\\
 FeI   & 7.52    &165&    5.72     &  0.12 &-1.80&& \\ 
 FeII   & 7.52    & 22&  5.84&0.12       &-1.68&&\\
 CoI   & 4.92     & 13 & 3.15  & 0.11&-1.77&0.03&\\ 
 NiI   & 6.23     &20 &4.32&0.41&-1.91&-0.12&\\
 NiII   & 6.23      & 1 &4.78&0.20&-1.46&0.34& \\
 CuI & 4.21      & 1& 1.60&0.20&-2.61&-0.81& \\
 ZnI & 4.62      & 2& 2.94&0.13&-1.68&0.11&\\
 SrII   & 2.92      &  2&1.04&0.56&-1.88&-0.08&\\
 YII    & 2.21      & 7 &0.66&0.05&-1.55&0.25&\\
 ZrII   & 2.62      & 3&1.24&0.07&-1.38&0.42&\\
 BaII  & 2.17      & 3&0.95&0.15&-1.23&0.57&\\
\hline
\end{tabular}
\tablefoot{LiI*  is for 3DNLTE. Tables  for the other  stars are provided on line.
The adopted solar abundances are from \citet[][]{Lodders2009}, except for N, O, and Fe, which are from \citet[][]{caffau2011SoPh..268..255C}, and Zr, which is from \citet[][]{caffau2011}. 
Li is from the Orgueil CI-chondrite \citep[][]{Lodders2009}.
}
\end{table}

\section{Discussion}\label{sec12}

The ratio between the Be and Li abundances could provide clues on the  nucleosynthetic  processes that produced these elements.
In the most extreme case of BPM\,3066  the Be enrichment comes along with  a Li overabundance. The ratio between the Be and the Li abundances in excesses over primordial is of about 8. This is  about the ratio  of the spallative cross sections of energetic CNO particles against protons and alphas at rest  thus providing a clear indication that both elements originate from spallation processes \citep{Steigman1992}.
A second case is that of HD~106038 that has  A($^9$Be) = 1.37 $\pm$ 0.12 \citep{tan2009} and  lithium A($^7$Li) = 2.49  \citep{Asplund2006}  or A($^7$Li) = 2.55   \citep{tan2009} and for which we derive a 3D NLTE abundance of A($^7$Li) = 2.59 $\pm$ 0.01 . This corresponds to a $^7$Li abundance $\sim$0.3 dex above the Spite plateau, and a $^9$Be enhancement of $\sim$1.2 dex relative to that of stars of similar metallicity. The resulting A(Li)/A(Be) excess ratio of $\sim$9 again confirms a spallative origin.
In the other stars, the Be abundance  is too small to produce an equivalent fraction of spallated  Li that can be detected in excess above the primordial level of A($^7$Li) $\approx 2.2$.

A beryllium abundance  of   A(Be) $\sim$0.5 dex would correspond to a  spallated lithium of   A(Li) = 1.4 dex which would produce an enhancement of only   $\sim$0.05 dex  above the primordial Li value, an excess  which is comparable to the typical Li abundance measurement errors and therefore  difficult to detect. However, as can be seen in Fig. \ref{fig:fig_libe}, the Li abundances in our stars lie on the upper envelope of the halo Li distribution, suggesting that some Li in addition to the primordial component may have been produced by spallation processes.
Thus, we have to rely on  the two most extreme cases  and  take them as representative for  the underlying process also in the remaining Be-rich stars, where the Li in excess to the primordial one  cannot be directly constrained.

\subsection{Stellar ages}

\begin{figure}
\includegraphics[width=0.5\textwidth]{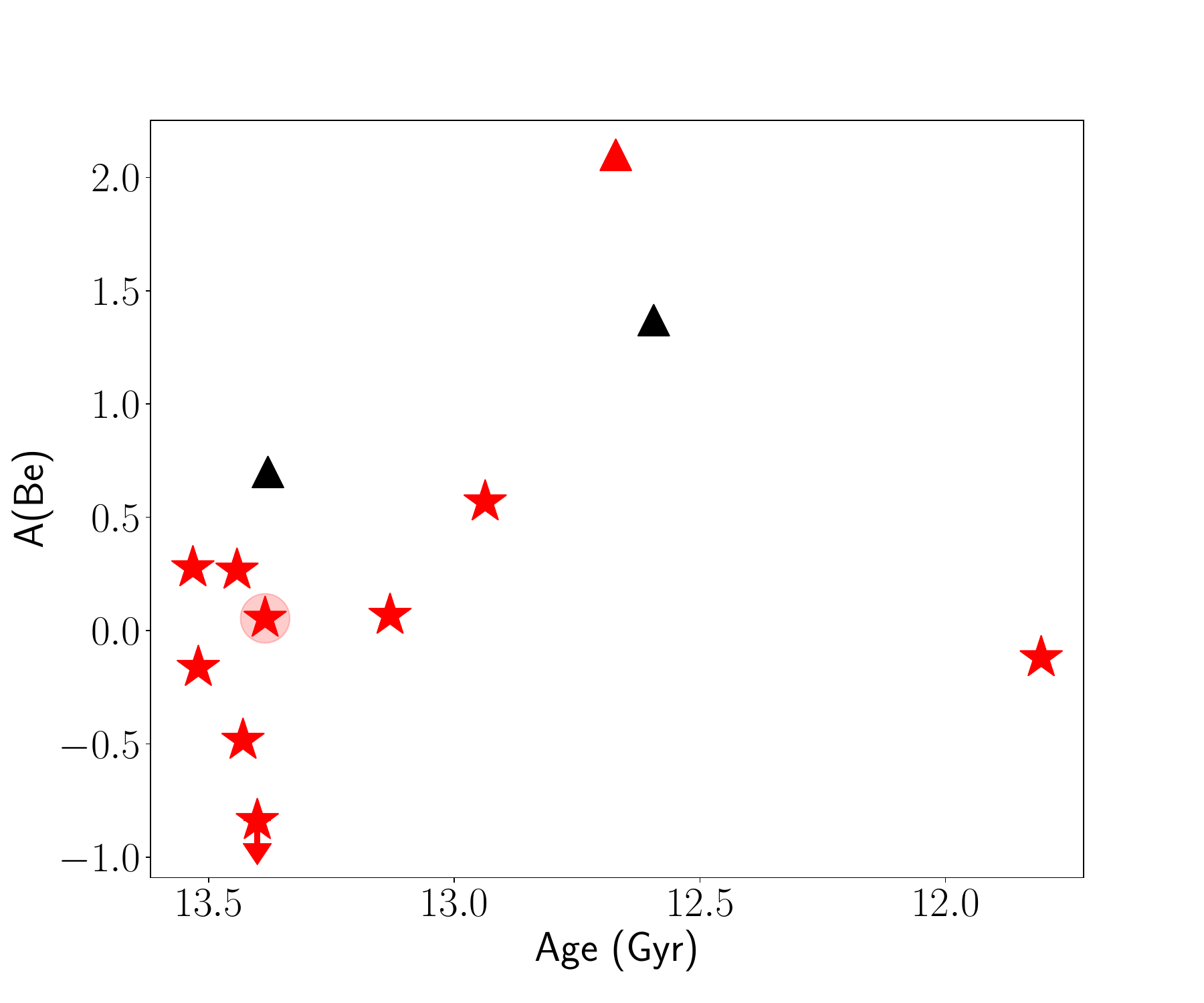}
   \caption{Beryllium abundances versus ages for  the program stars. The upper limit is for TS\,9, which is the star likely belonging to Thamnos-1. This is the level of beryllium expected to be present in  such old stars. Note that typical age uncertainties are of the order of about 1 Gyr }
    \label{fig:fig_ages}
\end{figure}

Using colour–magnitude diagram isochrone fitting, \citet{dodd2024arXiv240813763D} constrained the accretion times of different halo substructures. They found that Sequoia formed half of its stars by a lookback time of 12.0 Gyr, while Thamnos is on average slightly older, reaching the same stage by 12.3 Gyr. This suggests that Thamnos was accreted earlier than both GSE and Sequoia. However, the higher metallicity of Thamnos-2 compared to Sequoia seems at odds with this earlier accretion. A possible  explanation is that Thamnos-2 experienced rapid chemical enrichment soon after its formation—possibly driven by the same energetic events responsible for the observed Be over abundance.
 The  beryllium abundances versus stellar ages for the program stars are shown in Figure \ref{fig:fig_ages}. 
Stellar ages and masses for our sample were derived using the Python tool Stellar-Parameter-Inferred-Sistematically (SPInS; \citealt{LebretonReese2020,spins2020}) by combining observational data with the BASTI (Bag of Stellar Tracks and Isochrones; \citealt{Pietrinferni2021}) stellar evolution database, which provides models for extremely metal-poor stars.
 The stellar ages are remarkably old, spanning  an age between 11.8 and 13.5 Gyrs with errors in between 0.3 to 1 Gyr.
A significant scatter in Be abundances is apparent among the oldest stars. In very old stars, Be is expected to be near the minimum values observed, consistent with its smooth increase in the general Be–metallicity relation. The expected value is around A(Be) $\approx$ –1.0, or lower, as measured in TS\,9, which has a metallicity of [Fe/H] = –2.9 and is a likely member of Thamnos-1. Furthermore, the Be abundance does not appear to increase smoothly with age; in TS 5, the youngest star in the sample at 11.8 Gyr, the abundance remains roughly constant or even slightly decreases. This behavior strongly suggests prompt Be production, with varying degrees of dilution required to reproduce the observed scatter.

\subsection{Hypernova?}\label{sec124}

The detection of Be excess in a stellar group favours the hypothesis of hypernovae or other energetic events over that of planetary engulfment for the origin of such a chemical anomaly as it was proposed in \citep{Monaco2025A&A...697A..75M}. 
 
Hypernovae (HNe) are core-collapse supernovae with exceptionally large explosion energies, up to an order of magnitude higher than typical core-collapse SNe \citep{nomoto2013ARA&A..51..457N}. Their explosions accelerate CNO nuclei to energies $>$100 MeV, which then collide with the surrounding interstellar medium and fragment into lighter nuclei, efficiently producing $^9$Be and $^7$Li through spallation \citep{fields2002ApJ...581..389F}.
Note that \citet{fields2002ApJ...581..389F} calculated the yields of spallation products from HN explosions and predicted a $^7$Li/$^9$Be ratio of $\approx$8.6, very close to the values measured in BPM\,3066 and HD\,106038.

The absolute yields of Li and Be, however, are more uncertain. For example, model (b) for SN~1998bw in \citet{fields2002ApJ...581..389F} predicts $4.51\times10^{-5}\,M_\odot$ of $^7$Li and $6.71\times10^{-6},M_\odot$ of $^9$Be. Given that the observed Be overabundances correspond to (Be/H) mass fractions of order $3\times10^{-11}$, a  dilution of this material into $\sim 10^4\,M_\odot$ of interstellar medium is required to reproduce the observed levels of Li and Be. Note that 
the galaxy progenitor of Thamnos is estimated to have had a stellar mass of M$\star \le 5 \times 10^{6}\,M_\odot$ \citep{koppelman2019A&A...631L...9K}.
The same model also predicts ejecta masses of $\sim 2.9\,M_\odot$ of oxygen and $\sim 0.5\,M_\odot$ of iron.
When diluted into $\sim 10^4\,M_\odot$ of pristine interstellar medium, these yields result in [Fe/H], [O/H]  about  -1.3, -1.5, in good agreement with the abundances measured in our stars.

Several authors have noted a peculiar behaviour of the $\alpha$-elements in Thamnos candidate stars, 
which remain enhanced even at relatively high metallicities ([Fe/H]~$\gtrsim -1.5$; 
\citealt{koppelman2019A&A...631L...9K, dodd2023A&A...670L...2D, ceccarelli2025A&A...704A.180C}). 
Such a trend is not observed in stars of comparable metallicity in Local Group dwarf galaxies 
or in those accreted by the Milky Way. 
\citet{ceccarelli2025A&A...704A.180C} proposed that the high-$\alpha$ sequence originates from in-situ Galactic stars, 
estimating a contamination fraction of $\sim$78\%. 
However, \citet{dodd2023A&A...670L...2D} derived an upper limit of $\lesssim$ 50\% from 
chemical--dynamical modelling. 

An alternative  explanation we suggest here involves an energetic event that produced a prompt chemical enrichment 
while simultaneously reducing the gas mass available for dilution --- potentially the same event 
responsible for the observed beryllium production accounting for  the anomalous $\alpha$-element 
sequence.

\begin{figure}
\includegraphics[width=0.5\textwidth]{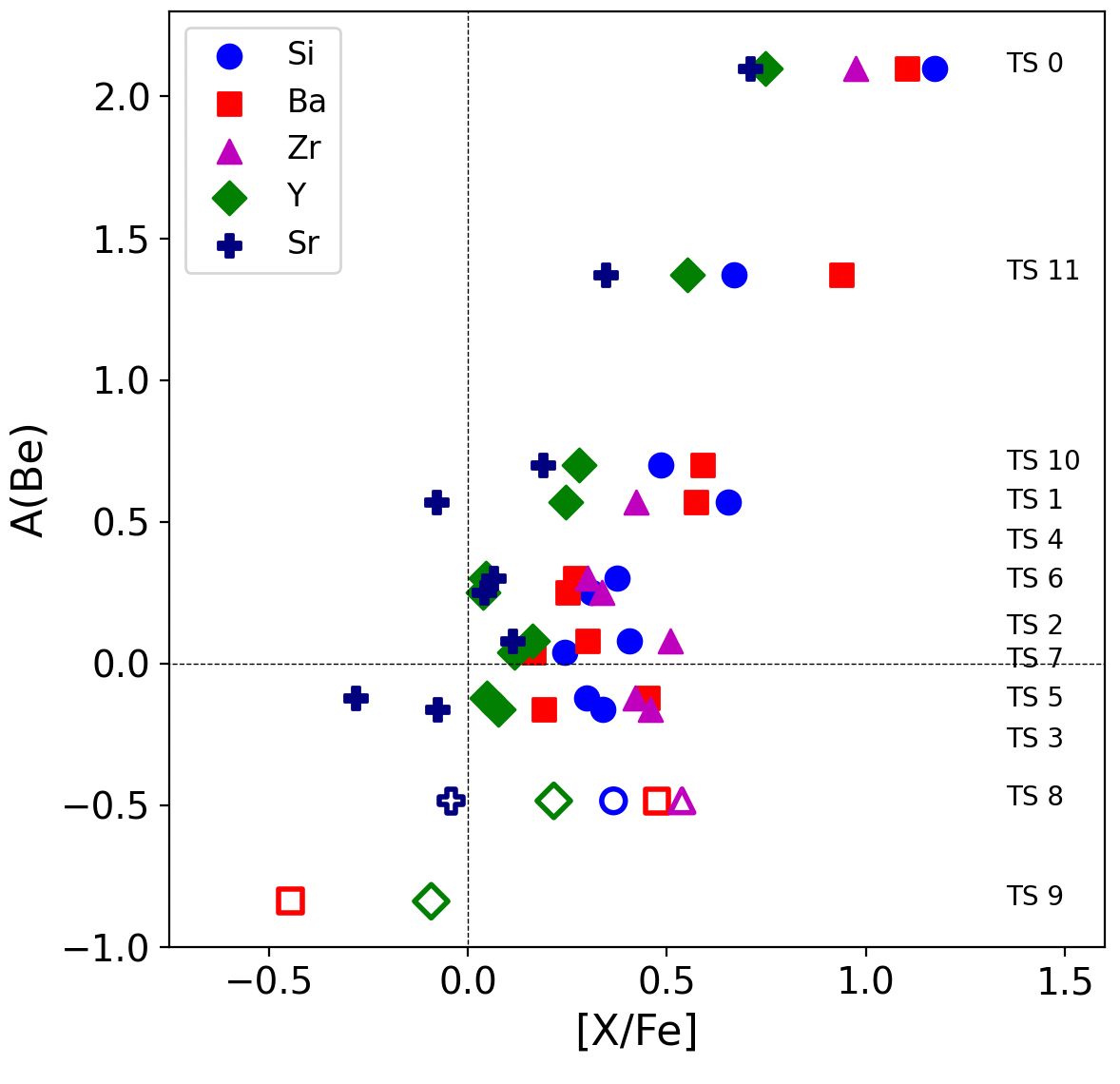}
    \caption{ 
    A(Be) is plotted against the [Si, Sr, Ba, Y, Zr/Fe] abundances. Empty symbols indicate the two Thamnos-1 stars. A strong correlation is found between the Be and Si abundances, spanning approximately 1.5 dex. A similar trend is observed for all the measured neutron-capture elements. }
    \label{fig:si}
\end{figure}

\begin{figure}
\includegraphics[width=0.5\textwidth]{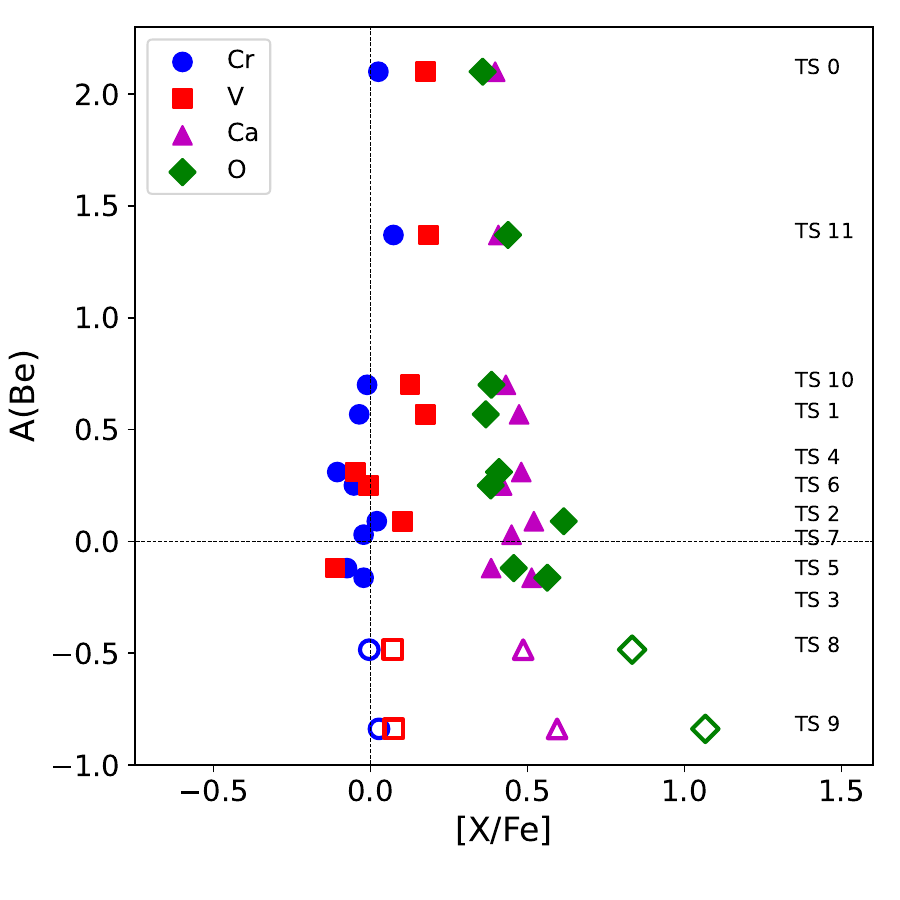}
    \caption{ A(Be) is plotted against the [Cr, V, Ca, O/Fe] abundances. These elements exhibit minimal variation, remaining constant within 0.2 dex regardless of the Be abundance. }
    \label{fig:cr}
\end{figure}

Thus, the varying degrees of Be enhancement can  be interpreted as the result of different dilution fractions. Stars with higher Be abundances formed from gas less mixed with pristine material, whereas those with lower Be reflect stronger dilution.
In hypernovae, oxygen burning proceeds more extensively at higher explosion energies, leading to greater production of burning products such as Si. As a result, hypernova nucleosynthesis is expected to produce high [Si/Fe] ratios \citep{nomoto2013ARA&A..51..457N}. If the observed Be excesses are indeed linked to reduced dilution, a positive correlation between Be abundance and [Si/Fe] should be observed. In Figure~\ref{fig:si}  the A(Be) abundance 
is plotted versus the [Si/Fe] derived from Table \ref{tab:tababb}. The values for BPM\,3066 from \citep{Monaco2025A&A...697A..75M}  and the HD 106038 and HD 132475 have been also included.  To note that also the n-capture elements show a peculiar enhancement.  In Figure~\ref{fig:si} the Ba and Si are also shown  and similar behaviour is observed with yttrium and zirconium. A  overabundance  of barium, yttrium and lanthanum was also pointed out  in \citet{ceccarelli2025A&A...704A.180C} in Thamnos candidate stars at similar metallicities. 
This correlation stands in striking contrast to other elements, which show abundances uncorrelated with the Be anomaly. In Figure~\ref{fig:cr}, Cr, V, Ca, and O are shown as typical examples of elements that do not correlate with Be. For these elements, the abundances are nearly uniform, with an intrinsic dispersion of less than 0.2 dex, whereas Be abundances span up to 1.5 dex
 Remarkably, oxygen does not correlate with beryllium, despite being one of the primary elements responsible for beryllium production through spallation. This is because oxygen is predicted to be depleted in HNe, where it is burned into silicon \citep{nomoto2013ARA&A..51..457N}. In fact, the only star in which the oxygen abundance is about 0.5 dex higher than that of calcium is TS,9, which is likely a member of the Thamnos-1 group and exhibits chemical evolution driven by CCSNe.

\subsection{Tidal interactions}

A possible way to produce cosmic rays, suitable to produce Li and Be
through spallation, that does not involve SN explosions, are tidal
shocks in interacting galaxies.
\citet{prodanovic2013} and \citet{prodanovic2019} investigated this
physical process, that they call Tidal Cosmic Rays (TCR) in the attempt to explain the
high $^6$Li abundance ($^6$Li/H
= $1.5 \times 10^{-10}$) tentatively measured by
\citet{howk2012} in the interstellar medium of the SMC. However, this value has since
been downward revised   by \citet{molaro2024}, who find it to be consistent with the Galactic value. Thus it is probably not necessary to resort to TCR to explain the SMC observations.
Yet this process may be relevant for gas-rich dwarf galaxies that are falling
in the Milky Way potential on plunging orbits \citep{prodanovic2013}.
In order for Li and Be produced by TCR in this case, to be observable
in stars, requires that these stars are formed in the starburst
triggered by the ram pressure exerted on the gas of the galaxy by the
Milky Way \citep[see e.g.][]{Wang2024}. After this star formation
episode the dwarf galaxy is stripped of its gas and can no longer form stars.
This scenario has some further appeal, Thamnos-2 may be simply be
the result  of the last starburst of Thamnos-1 as it was plunging
into the Galactic potential.
However, all of the stars in Thamnos-2
should have been formed in this last starburst of the incoming
dwarf galaxy, while we know that these starbursts form only a small
fraction of stars of the galaxy (less than 5\%).  Moreover, there is no 
mechanism that could slow down  the fast $^9$Be nuclei produced by the spallation to be incorporated into the forming stars,
rather than escaping the galaxy. Although we cannot rule out the possibility that 
TCR may be responsible for
the Be production in our program stars we consider that it is unlikely. 
Moreover, normal beryllium and lithium abundances are observed in other accreted systems, such as the GSE \citep{molaro2020}, further disfovouring this mechanism as the primary origin of the observed abundance patterns.

\section{Conclusions}\label{sec13}

We studied the abundances of Li and Be in a sample of ten stars, two of which likely belong  to Thamnos\,1 and eight  to Thamnos\,2.  The main conclusions of the work are summarized below:

\begin{itemize}
    \item { 
    
    We identified four new stars  of Thamnos-2 exhibiting beryllium overabundances clearly above the general Be–metallicity trend,  A fifth star, TS\,7, is potentially Be-rich, with A(Be) = 0.2, but its spectrum is quite noisy, making the abundance rather uncertain. Two additional stars, TS\,3 and TS\,5, lie above the general trend but still within 3$\sigma$, suggesting only moderate Be enhancement.On the other hand  two  stars of Thamnos-1, TS\,8 and TS\,9,    show Be abundances or limits consistent with typical Galactic Be behavior.
    }
     \item { 
     In addition, we found that the two other known Be-rich stars, HD 106038 and HD 132475,  share the same kinematical properties as Thamnos. This forms a consistent group of  Be-rich stars likely sharing a common origin. It is therefore possible that Be enhancement occurs only in Thamnos-2  providing an effective way to distinguish membership between these two structures.     }
      \item {
      In the two cases with the most extreme Be enhancements, we also observe Li abundances above the Spite plateau. The ratio between the Be and Li excesses, with respect to the  Galactic trends and Population II values, is approximately 8 — a characteristic signature of spallation processes.}
      
 \item {The production of light elements such as Be and Li could be associated with a hypernova or a similarly energetic event, in which energetic CNO nuclei collide with protons and alpha particles at rest in the interstellar medium. The varying degrees of Be enhancement observed among the stars could result from  different levels of dilution with the surrounding medium prior to star formation. The synthesis of Be could have been  accompanied by a prompt enrichment in iron, as evidenced by the fact that the stars with the highest Be abundances are also the most metal-rich in our sample. }
 
   \item {
        Stellar ages have been derived and are found to cluster around 13 Gyr. At this age, Be should be absent or very low, as observed in 
       the Thamnos -1 stars  TS\,9 and TS\,8. 
        In contrast, all the Thamnos-2 stars exhibit Be abundances that are too high for their ages, given that the Galactic enrichment of Be is a very slow process starting from essentially zero.   }

       \item { 
       Lithium abundances in stars  moderately Be-rich   are found to lie on, or slightly above, the Spite plateau. Considering that Thamnos-2 is a merger remnant, this result — together with similar findings for GSE — indicates that old stars originating outside the Milky Way share the same Li abundances as Galactic halo stars, and thus exhibit the same cosmological lithium problem. }

\end{itemize}

In summary, we identify a distinct population of Be-rich stars associated with Thamnos-2, likely originating from a single energetic event such as a hypernova. The combination of Be and Li abundances, stellar ages, and kinematics provides compelling evidence for this scenario and highlights the power of light-element abundances as tracers of Galactic halo substructures.

\begin{acknowledgements}
      Part of this work was supported by the Prin INAF 2022 (P.I P.M.).L. M. acknowledges support from ANID-FONDECYT Regular Project 1251809. GC acknowledges the financial
support under the National Recovery and Resilience Plan (NRRP), Mission 4,
Component 2, Investment 1.1, Call for tender No. 104 published on 2.2.2022
by the Italian Ministry of University and Research (MUR), funded by the European Union – NextGenerationEU – Project ‘Cosmic POT’ (PI: L. Magrini) Grant Assignment Decree No. 2022X4TM3H by the Italian Ministry of University and Research (MUR). G.C. acknowledges support from INAF through
the Large Grants EPOCH.
P.M and G.C acknowledge support from INAF through
the GTO GRANT  "Search for chemical fingerprints of the FIRST STARS in extremely metal poor CEMP-no stars).
Based on observations collected at the European Southern Observatory 
             under ESO programme [Programme ID  111.24HT (P.I. L. Monaco)
\end{acknowledgements}

\bibliography{thamnos}

\end{document}